\documentclass{ws-procs975x65}

\begin{document}


\wstoc{The Casimir effect in relativistic quantum field 
theories}{V.~M.~Mostepanenko}
\vspace*{-3cm}

\title{THE CASIMIR EFFECT IN RELATIVISTIC QUANTUM FIELD
THEORIES\footnote{This research has been partially supported by
DFG grant 436\,RUS\,113/789/0--2.}}

\author{V.~M.~MOSTEPANENKO\footnote{On leave from Noncommercial Partnership 
``Scientific Instruments'',
Tverskaya St. 11, Moscow, 103905, Russia.}${}^{,\ddag}$ }

\address{Center of Theoretical Studies Institute for Theoretical Physics, 
Leipzig University,\\
           Augustusplatz 10/11, 04109, Leipzig, Germany\\
${}^{\ddag}$E-mail: {Vladimir.Mostepanenko@itp.uni-leipzig.de}}
\noindent

\begin{abstract}
We review recent developments in the Casimir effect which arises 
in quantization volumes restricted by material boundaries and in
spaces with non-Euclidean topology. The starting point of our
discussion is the novel exact solution for the electromagnetic
Casimir force in the configuration of a cylinder above a plate.
The related work for the scalar Casimir effect in sphere-plate 
configuration is also considered, and the application region
of the proximity force theorem is discussed. Next we consider
new experiments on the measurement of the Casimir force between 
metals and between metal and semiconductor. The complicated 
problem connected with the theory of the thermal Casimir force
between real metals is analyzed in detail. The present situation 
regarding different theoretical approaches to the resolution of 
this problem is summarized. We conclude with new constraints on
non-Newtonian gravity obtained using the results of latest 
Casimir force measurements and compare them with constraints 
following from the most recent gravitational experiments.
\end{abstract}

\keywords{Casimir effect; exact solutions; Nernst heat theorem; 
non-Newtonian gravity. }

\bodymatter

\section{Introduction}

The Casimir effect\cite{1} is a particular type of vacuum polarization
which arises in quantization volumes restricted by material
boundaries and in spaces with non-Euclidean topology due to
distortions in the spectrum of zero-point oscillations of
relativistic quantized fields in comparison with the case
of free infinite Euclidean space-time. In case of volumes
restricted by material boundaries, the polarization energy
results in the Casimir force acting on these boundaries.
In spaces with non-Euclidean topology, the polarization
stress-energy tensor influences the geometry of space-time
through the Einstein equations of gravitational field. 
In both cases the applications of the Casimir effect are
extraordinary wide and range from condensed matter physics,
atomic physics and nanotechnology to gravitation and
cosmology (see monographs \refcite{2}--\refcite{5}
and reviews \refcite{6}--\refcite{8}).

During the last few years the Casimir force was measured with
increased precision in configurations
metal-metal\cite{9}${}^{-}$\cite{19} and
metal-semiconductor.\cite{20}${}^{-}$\cite{22}
The theory of the Casimir effect was widened to incorporate 
real material properties\cite{7} and more complicated
geometrical configurations.\cite{23,24} Much attention was
given to the controversial problem of the thermal Casimir
force between real metals (see discussion in
Refs.~\refcite{25,26}) and
dielectrics.\cite{27}${}^{-}$\cite{29}
The results of precise measurements of the Casimir force
between metal surfaces were used for obtaining stronger
constraints on the Yukawa-type corrections to
Newtonian gravitational law predicted in unified gauge
theories, supersymmetry and
supergravity.\cite{17}${}^{-}$\cite{19,30}

In the present paper we discuss the above most important
achievements in the Casimir physics for the period after
the Xth Marcel Grossmann Meeting which was hold in
July, 2003 at Rio de Janeiro. In our opinion, the
theoretical achievement of major significance during
this period is the obtaining the exact solution for the
electromagnetic Casimir force in configuration of a
cylinder above a plate\cite{23} (see also further
development of this matter in Ref.~\refcite{24}). 
The related work was done for the scalar Casimir force in
configurations of a sphere or a cylinder above a
plate.\cite{24,31}
In Ref.~\refcite{32} the scalar Casimir effect in the same
configurations was considered numerically using the
worldline algorithms. The combination of the exact
analytical and precise numerical methods permitted
to make some conclusions on the validity limits
of the so-called proximity-force theorem (PFT) which
is heavily used in the experimental investigation
of the Casimir force. All these results are discussed
in Sec.~2 of the present paper.

Sec.~3 is devoted to new experiments on the measurement of
the Casimir force between metals\cite{17}${}^{-}$\cite{19}
and between metal and
semiconductor.\cite{20}${}^{-}$\cite{22}
The experiments\cite{17}${}^{-}$\cite{19} using a
micromechanical torsional oscillator permitted for the
first time to achieve the recordly low total experimental
error of about 0.5\% within a wide separation range
and reliably decide between different competing approaches
to the theoretical description of the thermal Casimir
force. The experiments\cite{20}${}^{-}$\cite{22}
using an atomic force microscope opened new
prospective opportunities
for the control of the Casimir force in nanodevices by
changing the charge carrier density in a semiconductor
test body.

The complicated theoretical problems related to the
thermal Casimir force are discussed in Sec.~4. As underlined 
in this section, the theoretical approach proposed by some
authors for real metals\cite{26} is not only in
contradiction to experiment, but inavoidably results
in a violation of the Nernst heat theorem.\cite{25,33,34}
What is more, we stress that the same problems, as for
real metals, arise for the Casimir force in
configurations of two dielectrics and metal-dielectric
if dc conductivity of a dielectric plate is taken into
account.\cite{27}${}^{-}$\cite{29} This suggests that there 
are serious restrictions in a literal application of the
Lifshitz theory to real materials. Some phenomenological
approaches on how to avoid contradictions with
theormodynamics and experiment proposed in literature
are discussed.

In Sec.~5 reader finds the review of new constraints on
non-Newtonian gravity obtained from recent measurements
of the Casimir force between metallic test
bodies.\cite{17}${}^{-}$\cite{19,30} These constraints
are compared with those obtained from gravitational
measurements.

Sec.~6 contains our conclusions and discussions.

\section{New exact solutions in configurations with curved
boundaries}

It is common knowledge that Casimir\cite{1} found the
exact expression 
\begin{equation}
F(z)=-\frac{\pi^2}{240}\,\frac{\hbar c}{z^4}
\label{eq1}
\end{equation}
\noindent
for the fluctuation force 
of electromagnetic origin per unit area acting between two
plane-parallel ideal metal plates at a separation $z$.
Lifshitz theory\cite{35} generalized Eq.~(\ref{eq1}) for
the case of two parallel plates described by a 
frequency-dependent dielectric permittivity 
$\varepsilon(\omega)$. Experimentally it is hard to
maintain the parallelity of the plates. Because of this,
most of experiments were performed using  the configuration
of a sphere above a plate. The configuration of a cylinder
above a plate also presents some advantages if to compare
with the case of two parallel plates. Unfortunately,
over many years it was not possible to obtain exact
expressions for the Casimir force in these configurations.
For this reason, the approximative proximity-force
theorem\cite{36} (PFT) was used to compare experiment with
theory. According to the PFT, at short separations
($z\ll R$) the Casimir forces between an ideal metal cylinder
(per unit length) or a sphere and a plate are given by
\begin{equation}
F_c(z)=-\frac{\pi^3}{384\sqrt{2}}\,\sqrt{\frac{R}{z}}\,
\frac{\hbar c}{z^3},
\qquad
F_s(z)=-\frac{\pi^3}{360}\,\frac{\hbar c R}{z^3},
\label{eq2}
\end{equation}
\noindent
where $R$ is a sphere or a cylinder radius.

Within the PFT it is not possible to control the error
of the approximate expressions in 
Eq.~(\ref{eq2}). From dimensional considerations it
was evident\cite{7} that the relative error in Eq.~(\ref{eq2})
should be of order of $z/R$, but the numerical coefficient
near this ratio remained unknown. In fact, rigorous
determination of the error, introduced by the application of 
the PFT, requires a comparison of Eq.~(\ref{eq2}) with the
exact analytical results or with precise numerical
computations in respective configurations. One such result for
the electromagnetic Casimir effect was first obtained\cite{23}
for a cylinder above a plate using a path-integral
representation for the effective action. Eventually,
the Casimir energy is expressed through the functional 
determinants of infinite matrices with elements given in terms
of Bessel functions.\cite{23} The analytic asymptotic behavior
of the exact Casimir energy at short separations was found in
Ref.~\refcite{24}. It results in the following expression for
the Casimir force at $z\ll R$:
\begin{equation}
F_c(z)=-\frac{\pi^3}{384\sqrt{2}}\,\sqrt{\frac{R}{z}}\,
\frac{\hbar c}{z^3}\left[1-\frac{3}{5}\,
\left(\frac{20}{3\pi^2}-\frac{7}{36}\right)\,
\frac{z}{R}\right].
\label{eq3}
\end{equation}
\noindent
Eq.~(\ref{eq3}) is of much importance. It demonstrates that
the relative error of the electromagnetic Casimir force between
a cylinder and a plate calculated using the PFT is equal to
$-0.288618z/R$. Thus, for typical parameters of $R=100\,\mu$m
and $z=100\,$nm this error is approximately equal to
only 0.03\%.

For a sphere above a plate the analytic solution in the
electromagnetic case is not yet obtained. The scalar Casimir
energy for a sphere above a plate is found in Refs.~\refcite{24}
and \refcite{31}. However, the asymptotic expression at short
separations similar to Eq.~(\ref{eq3})
is not found. In Ref.~\refcite{32} the scalar Casimir energies
for both a sphere and a cylinder above a plate are computed
numerically using the worldline algorithms. It was supposed
that a scalar field satisfies Dirichlet boundary conditions.
As was noticed in Ref.~\refcite{32}, the Casimir energies
for the Dirichlet scalar should not be taken as an estimate
for those in electromagnetic case. In addition, it should be
stressed that the errors of the PFT calculated in
Ref.~\refcite{32} are related to the Casimir energy and not to
the experimentally measured Casimir force. This makes all errors
larger. To illustrate, if we were considering the error of
the PFT in application to the electromagnetic Casimir energy
between a plate and a cylinder [instead of the force  
considered in Eq.~(\ref{eq3})], the value of $-0.48103z/R$ would be 
obtained  as a negative error of the PFT.\cite{24} 
The magnitude of the latter is by a factor 
of 1.6667 larger than the error obtained above for a force.

Eq.~(\ref{eq3}) confirms that PFT works well at short 
separations and reproduces the exact result with a very high
precision. This justifies the use of the PFT for the
interpretation of experimental data.
In Refs.~\refcite{37,38} it was claimed, however, that in
the configurations of sinusoidally corrugated plates or
a sphere above a plate the PFT overestimates the lateral
Casimir force by up to 30--40\%. Comment~\refcite{39}
demonstrates that these claims are not warranted.
In Refs.~\refcite{37,38} metal is described by the
plasma model with a plasma wavelength $\lambda_p=136\,$nm
for Au. Deviations of the ``exact'' results 
obtained in Refs.~\refcite{37,38} from those given by
the PFT in plate-plate configuration are presented in
Fig.~1 of Ref.~\refcite{37} (Fig.~11 of Ref.~\refcite{38})
in terms of function $\rho$ versus $k=2\pi/\lambda_C$,
where $\lambda_C$ is the corrugation wavelength.
According to this figure, the lateral force amplitude
is less by 16\% than the value given by the PFT
for configuration with $\lambda_C=1.2\,\mu$m and plate
separation $z=200\,$nm (it is supposed that corrugation
amplitudes are much less than $z,\>\lambda_p$
and $\lambda_C$). This result of
Refs.~\refcite{37,38} is in contradiction with a more
fundamental path-integral theory formulated for ideal
metals.\cite{40} It is easily seen, that the quantity
$\rho$, plotted in the above-mensioned figures as a function
of $k$ at different $z$ and with a fixed $\lambda_p$, 
is, in fact, a function of
$kz$. Thus, for corrugated plates with rescaled
$\lambda_C=12\,\mu$m and $z=2\,\mu$m (but with the same
$kz$) the deviation of the lateral force amplitude from
the PFT value is still 16\%. At $z=2\,\mu$m, however, the
nonideality of a metal does not play any important role, 
and Ref.~\refcite{40} demonstrates the complete agreement 
between the exact result and the PFT if (as it holds in our
case) $z$ is several times less than $\lambda_C$. Note that if
the latter condition is not satisfied, the PFT underestimates
the force amplitude,\cite{40} and not overestimates it as
in Refs.~\refcite{37,38}.

In the Reply\cite{39a} to the Comment\cite{39} the authors
of Refs.~\refcite{38,39} claim that the above arguments
raising doubts on their predictions are based on a mistake. 
This claim is in error. Reference~\refcite{39a} is
right that generally the case of perfectly reflecting mirrors
is recovered in the limit $\lambda_p\to 0$. In the
formalism of Refs.~\refcite{37,38}, however, this limiting
transition is forbidden by the condition that the corrugation
amplitudes are much less than $\lambda_p$. Thus, for fixed
corrugation amplitudes the limiting case of ideal metal cannot
be achieved by decreasing $\lambda_p$. On the contrary, the 
formalism of Refs.~\refcite{37,38} allows any increase of 
$\lambda_C$ and $z$, and this was used in the Comment\cite{39}.
At separations $z\gg\lambda_p$ (in the Comment $z=2\,\mu$m)
real metal behaves like ideal metal and all results should coincide
with those for ideal metals as obtained in the path-integral
approach.\cite{40,41} In fact, however, the results of
Refs.~\refcite{37,38} at large separations are drastically
different from the results of Refs.~\refcite{40,41} obtained 
for ideal metals. This leads us to the conclusion that the
predictions of Refs.~\refcite{37,38} made for the
configuration of two corrugated plates are in error.

For the experimental configuration of a sphere above a
plate Refs.~\refcite{37,38} obtain the ``exact''
computational value 0.20\,pN for the amplitude of the
lateral force at a separation $z=221\,$nm between the test 
bodies with corrugation amplitudes equal to
$A_1=59\,$nm and $A_2=8\,$nm. According to Refs.~\refcite{37,38},
the linear in the corrugation amplitudes version of the PFT
gives instead 0.28\,pN, i.e., 40\% difference. At this
point Ref.~\refcite{39} stresses that the amplitudes
considered are not small comparing to $z$ (for instance,
$A_1/z=0.27$) and another assumption
$A_1,\,A_2\ll\lambda_p$ used in Refs.~\refcite{37,38}
is also violated (for instance, $A_1/\lambda_p=0.43$).
It is not surprising, then, that Refs.~\refcite{37,38}
arrive at a force amplitude of 0.20\,pN so far away from
the value of 0.33\,pN obtained theoretically using the
complete PFT and that of $0.32\pm 0.077\,$pN measured
experimentally at 95\% confidence in Ref.~\refcite{16}. Thus,
the approach used in Refs.~\refcite{37,38} is not only
in contradiction with a more fundamental path-integral
theory\cite{40} but is also excluded by experiment.\cite{15}
This justifies the conclusion of Refs.~\refcite{40,41}
that the PFT presents correct description of the lateral
Casimir force between corrugated test bodies when the
separation is several times less than the corrugation period.

\section{New precise measurements of the Casimir force
between metal and semiconductor test bodies}

The most important experiments on the Casimir force after the
Xth Marcel Grossmann Meeting were performed at Purdue University
--- Indiana State University\cite{17}${}^{-}$\cite{19} and at
the University of California, Riverside.\cite{20}${}^{-}$\cite{22}
Two experiments in Refs.~\refcite{17}--\refcite{19} are devoted
to the determination of the Casimir pressure between two 
Au-coated plates using the dynamic techniques based on a
micromechanical torsional oscillator. The improved version of
this experiment is described in Refs.~\refcite{18,19}. The two
test bodies of the micromechanical oscillator are a sphere
and a plane plate. Sphere is oscillating with the angular
resonant frequency, and the shift of this frequency under the
influence of the Casimir force $F$ acting between a sphere
and a plate was measured as a function of separation $z$ in
the region from 160 to 750\,nm. From this shift one can
find\cite{14a,17}${}^{-}$\cite{19} the force gradient
$\partial F/\partial z$ and using the PFT arrive to the
equivalent Casimir pressure
\begin{equation}
P(z)=-\frac{1}{2\pi R}\,\frac{\partial F(z)}{\partial z}.
\label{eq4}
\end{equation}
\noindent
This experiment is characterized by a very low total
experimental error which was determined at a 95\% confidence
level and varies between 0.55 and 0.60\% in a wide separation
region from 170 to 350\,nm.

The obtained experimental results were compared with
different theoretical approaches using the Lifshitz
theory\cite{35} and tabulated optical data for the
complex index of refraction.\cite{42}
In this comparison all corrections due to surface
roughness, nonzero temperature, sample-to-sample
variations of optical data, errors of the PFT, effects of
spatial nonlocality and of patch potentials were carefully
analyzed and taken into account in a conservative way.
Specifically, the error of the PFT was conservatively estimated
as equal to $z/R$, whereas recent results presented in Sec.~2
lead to several times smaller error. It was concluded that data 
are consistent with the surface impedance approach 
to the thermal Casimir force at the
laboratory temperature $T=300\,$K (see Fig.~1, left, where
the differences between theoretical, $P^{\rm th}$, and
experimental, $P^{\rm exp}$, Casimir pressures are plotted
versus separation).
The data were found to be consistent also 
with the theoretical approach using the
plasma model at $T=300\,$K, and with the theoretical computations 
at zero temperature. At the same time, Fig.~1, right, shows that 
experimental data exclude theoretical Casimir pressures,
$\tilde{P}^{\rm th}$, computed using the Drude model 
approach at $T=300\,$K
(discussion of different theoretical approaches is contained in
Sec.~4).
\begin{figure}[t]
\vspace*{-7cm}
\hspace*{-2.5cm}
\psfig{file=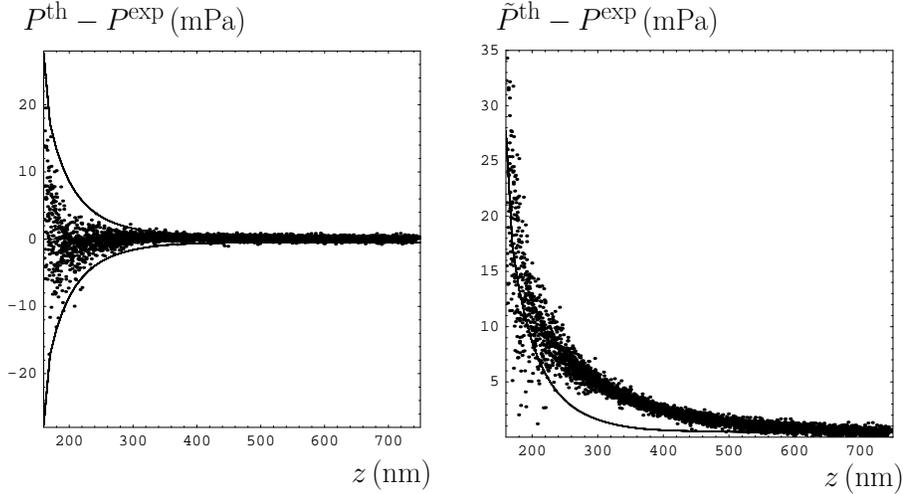,width=8in}
\vspace*{-16cm}
\caption{Differences of the theoretical and experimental Casimir 
pressures versus separation (dots) and the 95\% confidence
intervals (solid lines). The theoretical pressures $P^{\rm th}$ 
(left figure) are computed using the impedance approach
and $\tilde{P}^{\rm th}$ (right figure) using the Drude model approach.}
\end{figure}

The experiment by using a micromechanical torsional oscillator
has permitted also to obtain stronger constraints on
non-Newtonian gravity which are considered in Sec.~5.

Three experiments in Refs.~\refcite{20}--\refcite{22} are devoted
to the measurement of the Casimir force acting between Au-coated
sphere and single-crystal Si plates with different charge carrier 
densities using an atomic force microscope. In Ref.~\refcite{20} 
B-doped Si plate with a resistivity $\rho\approx 0.0035\,\Omega\,$cm
and concentration of charge carriers
$n\approx 3\times10^{19}\,\mbox{cm}^{-3}$ was used.
The measured force-distance relation of the Casimir force was
compared with two theoretical dependences. One of them was
computed for this sample and another one for a sample made of Si
with high resistivity equal to $1000\,\Omega\,$cm.
It was found that theoretical results computed for the
semiconductor plate used in experiment are consistent with
the data. At the same time, theoretical results computed for 
high-resistivity Si are experimentally excluded at 70\% confidence. 
This suggests that the Casimir force is sensitive to the
conductivity properties of semiconductors.
\begin{figure}[t]
\vspace*{-7cm}
\hspace*{-2cm}
\psfig{file=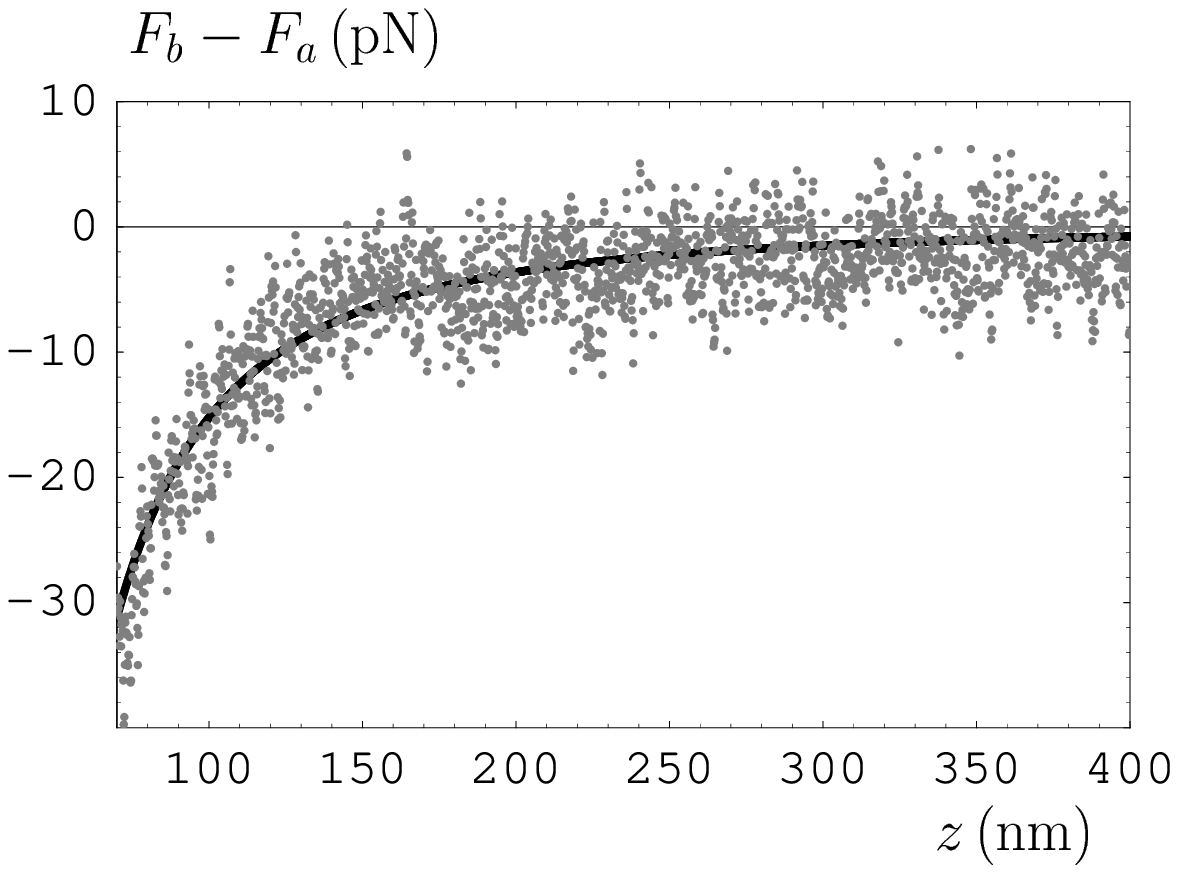,width=6in}
\vspace*{-8.5cm}
\caption{The differences of the mean measured Casimir forces
of the lower and higher resistivity Si (dots) and respective
theoretical difference (solid line) versus separation}
\end{figure}

The obtained results were confirmed in the direct measurement of
the difference Casimir force acting between Au-coated sphere and
two P-doped Si plates of different charge carrier
densities.\cite{21} One of the silicon plates (sample $a$)
had the resistivity $\rho_a\approx 0.43\,\Omega\,$cm
and the concentration of charge carriers
$n_a\approx 1.2\times10^{16}\,\mbox{cm}^{-3}$.
Another one (sample $b$) had much lower  
resistivity $\rho_b\approx 6.4\times 10^{-4}\,\Omega\,$cm
and much higher concentration of charge carriers
$n_b\approx 3.2\times10^{20}\,\mbox{cm}^{-3}$.
In Fig.~2, taken from Ref.~\refcite{21}, the difference
of experimental mean Casimir forces, acting between Au-coated
sphere and samples $b$ and $a$, $F_b-F_a$, versus separation
is shown as dots. The theoretically calculated differences
using the Lifshitz formula are shown by the solid line.
Within the separations from 70 to 100\,nm the mean difference
in the measured Casimir forces exceeds the experimental error
of force difference. This permits a conclusion that in
Ref.~\refcite{21} the influence of charge carrier density
of a semiconductor on the Casimir force was experimentally
measured for the first time.

The third experiment on the measurement of the Casimir
force between Au-coated sphere and single-crystal Si plate
demonstrates a new physical phenomenon, the modulation
of the Casimir force with laser light.\cite{22}
In the absence of light the used Si plate had a relatively
high resistivity $\rho\approx 10\,\Omega\,$cm
and relatively low concentration of charge carriers
$n\approx 5\times10^{14}\,\mbox{cm}^{-3}$.
This plate was illuminated with 514\,nm pulses,
obtained from an Ar laser. In the presence of pulse
the concentration of charge carriers increases up to
$n\approx 2\times10^{19}\,\mbox{cm}^{-3}$.
The difference of the Casimir forces in the presence and
in the absence of pulse, $\Delta F$, was measured using
an atomic force microscope within the separation range
from 100 to 500\,nm. The experimental results\cite{22}
are shown in Fig.~3 as dots versus separation. 
In the same 
figure the solid line is computed using the Lifshitz formula
under the assumption that in the absence of laser light Si
possesses a finite static dielectric permittivity
$\varepsilon^{Si}(0)=11.66$. The dashed line is computed
taking into account the dc conductivity of Si in the absence
of laser light at frequencies much below the first
Matsubara frequency. As is seen in Fig.~3, the solid
line is in excellent agreement with the experimental data,
whereas the dashed line is in disagreement with data.
Physical consequences following from this observation
are discussed in the next section.
\begin{figure}[t]
\vspace*{-7cm}
\hspace*{-2cm}
\psfig{file=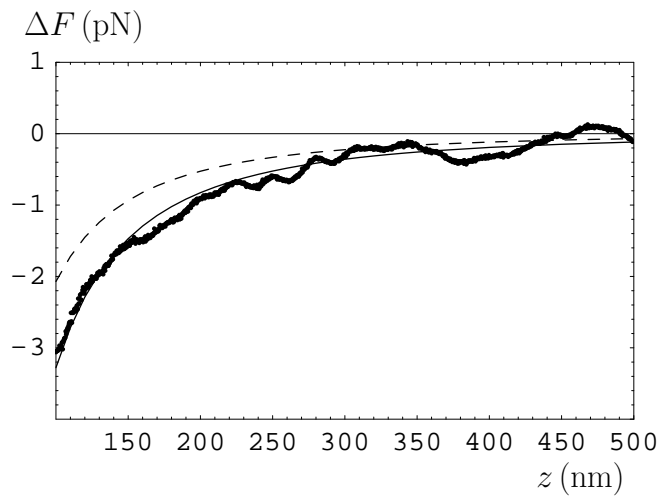,width=6in}
\vspace*{-8cm}
\caption{The differences of the mean measured Casimir forces
with laser pulse on and off (dots) versus separation.
The respective theoretical differences are computed
under the assumption of finite static dielectric
permittivity of Si in the absence of laser light (solid line) 
and taking dc conductivity of high-resistivity Si into account 
(dashed line).}
\end{figure}

The demonstrated dependence of the Casimir force between a metal 
and a semiconductor on the density of charge carriers in
semiconductor can be applied in nanodevices of the next
generations such as micromirrors, nanotweezers and nanoscale
actuators. In so doing, the density of charge carriers can be
changed either by doping and/or due to irradiation of a device
by laser light leading to respective variations in the
magnitude of the Casimir force.

Since the Xth Marcel Grossmann Meeting in 2003, some other experiments
on the Casimir force have been proposed. One could mention
the proposal to measure the influence of the Casimir energy on the
value of the critical magnetic field in superconductor phase
transitions, \cite{43} the suggestion to measure the Casimir
torques using the repulsive force due to liquid layers,\cite{44}
and the proposed Casimir force measurements at large 
separations.\cite{45}${}^{-}$\cite{46a} 
Special attention was attracted to new
techniques for the measurement of the Casimir force.
Thus, in Ref.~\refcite{47} the holographic interferometer
was first applied for optical detection of mechanical
deformation of a macroscopic object induced by the Casimir
force. All this demonstrates that there are considerable
opportunities in the experimental investigation of the
Casimir force and in applications of the Casimir effect.

\section{Problems in the theory of thermal Casimir
force between metals and dielectrics}

During all the period between the Xth and XIth Marcel Grossmann
Meetings the problem of the thermal Casimir force was hotly
debated. Until 2005, only the case of two plates made of real
metal was the subject of controversy. In 2005 it was shown,
however, that the case of two dielectric plates leads to problems
as well.\cite{27}

We start from the Lifshitz formula for the free energy of the
van der Waals (Casimir) interaction between two semispaces with
a gap of width $z$ in thermal equilibrium at temperature $T$:
\begin{eqnarray}
&&
{\cal F}(z,T)=\frac{k_BT}{2\pi}\sum\limits_{l=0}^{\infty}
\left(1-\frac{1}{2}\delta_{l0}\right)
\int\limits_{0}^{\infty}k_{\bot}\,dk_{\bot}
\label{eq5} \\
&&
\phantom{aaaaa}
\times\left\{\ln\left[1-r_{\rm TM}^2(\xi_l,k_{\bot})e^{-2q_lz}\right]
+\ln\left[1-r_{\rm TE}^2(\xi_l,k_{\bot})e^{-2q_lz}\right]\right\}.
\nonumber
\end{eqnarray}
\noindent
Here $k_B$ is the Boltzmann constant, 
$\xi_l=2\pi k_B Tl/\hbar$ are the Matsubara frequencies,
$q_l=(k_{\bot}^2+\xi_l^2/c^2)^{1/2}$, $k_{\bot}$ is the
projection of the wave vector on the boundary planes of 
semispaces, and $r_{\rm TM,TE}(\xi_l,k_{\bot})$ are the
reflection coefficients for two independent polarizations
of the electromagnetic field (transverse magnetic and
transverse electric modes).

In the original formulation of the Lifshitz theory the semispace
material is described by using the approximation of dielectric
permittivity $\varepsilon(\omega)$ depending only on the
frequency, and the continuity conditions
\begin{equation}
E_{1t}=E_{2t}, \quad
B_{1t}=B_{2t}, \quad
D_{1n}=D_{2n}, \quad
B_{1n}=B_{2n}
\label{eq6}
\end{equation}
\noindent
for the electric field, magnetic induction and electric
displacement on boundary planes. Thus, the Lifshitz theory
does not take into account the effects of spatial
dispersion. In this model case the reflection coefficients
take the form
\begin{equation}
r_{\rm TM}(\xi_l,k_{\bot})=\frac{\varepsilon_lq_l-
k_l}{\varepsilon_lq_l+k_l},
\qquad
r_{\rm TE}(\xi_l,k_{\bot})=\frac{k_l-q_l}{k_l+q_l},
\label{eq7}
\end{equation}
\noindent
where $k_l=\sqrt{k_{\bot}^2+\varepsilon_l\xi_l^2/c^2}$
and $\varepsilon_l=\varepsilon(i\xi_l)$.

The central point of the debates is the term of Eq.~(\ref{eq5}) with
$l=0$ (the so-called zero-frequency term). At large separations
(high temperatures) it is dominant, and all terms with $l\geq 1$
are negligibly small independently of the specific form of
$\varepsilon(\omega)$. The case of ideal metal plates is
obtained from Eqs.~(\ref{eq5}), (\ref{eq7}) using the so-called
Schwinger prescription,\cite{5,48} i.e., that one should take
limit $\varepsilon\to\infty$ first and set $l=0$ afterwards.
Using this prescription, for ideal metal plates one obtains
\begin{equation}
r_{\rm TM}(0,k_{\bot})=1,
\qquad
r_{\rm TE}(0,k_{\bot})=1.
\label{eq8}
\end{equation}
\noindent
The same result follows for ideal metal independently of the
Lifshitz formula from thermal quantum field theory with
boundary conditions in the Matsubara formulation. Thus, at
large separations (in fact at separations larger 
than $6\,\mu$m at $T=300\,$K) it follows
\begin{equation}
{\cal F}(z,T)=-\frac{k_BT}{8\pi z^2}\,\zeta(3),
\label{eq9}
\end{equation}
\noindent
where $\zeta(3)$ is the Riemann zeta function.
Notice that Eq.~(\ref{eq9}) is in agreement with
the classical limit based on the Kirchhoff's law.\cite{49,50a}

Refs.~\refcite{50}--\refcite{53} (see also Ref.~\refcite{26})
suggested to calculate the thermal Casimir force
by describing the properties of real metals at low
frequencies via the dielectric permittivity of the Drude
model
\begin{equation}
\varepsilon(i\xi)=1+
\frac{\omega_p^2}{\xi\left[\xi+\gamma(T)\right]},
\label{eq10}
\end{equation}
\noindent
where $\omega_p$ is the plasma frequency and $\gamma(T)$ is
the relaxation parameter. Substituting Eq.~(\ref{eq10}) in
Eq.~(\ref{eq7}) we obtain
\begin{equation}
r_{\rm TM}(0,k_{\bot})=1,
\qquad
r_{\rm TE}(0,k_{\bot})=0.
\label{eq11}
\end{equation}
\noindent
Eq.~(\ref{eq11}) is preserved also in the limit of ideal metal
plates, and is thus in contradiction with Eq.~(\ref{eq8}).
From Eqs.~(\ref{eq5}) and (\ref{eq11}) at large separations
one arrives at the result
\begin{equation}
{\cal F}(z,T)=-\frac{k_BT}{16\pi z^2}\,\zeta(3)
\label{eq12}
\end{equation}
\noindent
instead of Eq.~(\ref{eq9}). This result is in contradiction
with the classical limit.

Real metals in the frequency region of infrared optics are
well described by the dielectric permittivity of the plasma
model
\begin{equation}
\varepsilon(i\xi)=1+\frac{\omega_p^2}{\xi^2}.
\label{eq13}
\end{equation}
\noindent
If one estrapolates this model to low frequencies, the reflection
coefficients become\cite{54,55}
\begin{equation}
r_{\rm TM}(0,k_{\bot})=1,
\qquad
r_{\rm TE}(0,k_{\bot})=\frac{\sqrt{c^2k_{\bot}^2+\omega_p^2}-
k_{\bot}}{\sqrt{c^2k_{\bot}^2+\omega_p^2}+k_{\bot}}.
\label{eq14}
\end{equation}
\noindent
In the limiting case of ideal metal plates 
it holds $\omega_p\to\infty$ and Eq.~(\ref{eq14}) agrees with
Eq.~(\ref{eq8}) because $r_{\rm TE}(0,k_{\bot})\to 1$.
At large separations the plasma model leads to Eq.~(\ref{eq9})
in agreement with the classical limit.

It is notable that the plasma model predicts small thermal
corrections to the Casimir force at short
separations in qualitative agreement with the case of
ideal metals (a fraction of a percent at separations below
$1\,\mu$m). Much larger thermal corrections at short
separations are predicted by using the Drude model
(19\% of the force at $z=1\,\mu$m).

As was mentioned above, the dielectric permittivity depending
on the frequency provides only an approximative description of
metals because it disregards the effects of spatial dispersion. 
Another approximative description of metals is provided by
the Leontovich impedance boundary condition
\begin{equation}
\mbox{\boldmath{$E$}}_t=Z(\omega)
\left[\mbox{\boldmath{$B$}}_t\times\mbox{\boldmath{$n$}}\right],
\label{eq15}
\end{equation}
\noindent
where the index $t$ labels the field components tangential to
the plates, {\boldmath{$n$}} is the unit vector directed into
the medium, and impedance function $Z(\omega)$ is found
from the solution of kinetic equations.\cite{56}
It is notable that the Leontovich impedance is well defined
even in some frequency regions (for example, in the region
of the anomalous skin effect in which the spatial dispersion is present)
where the description in terms of $\varepsilon(\omega)$ is not
possible. At the same time, the Leontovich impedance is
not applicable at separations $z<\lambda_p=2\pi c/\omega_p$,
where the inequality $Z\ll 1$ may be violated and the boundary
condition (\ref{eq15}) cannot be used.
In the frequency regions where both quantities are well
defined it holds $Z(\omega)=1/\sqrt{\varepsilon(\omega)}$.

In terms of Leontovich impedance, the reflection
coefficients in the Lifshitz formula take the
form\cite{57,58}
\begin{equation}
r_{\rm TM}(0,k_{\bot})=\frac{cq_l-Z_l\xi_l}{cq_l+Z_l\xi_l},
\qquad
r_{\rm TE}(0,k_{\bot})=\frac{\xi_l-cq_lZ_l}{\xi_l+cq_l Z_l},   
\label{eq16}
\end{equation}
\noindent
where $Z_l=Z(i\xi_l)$. The zero-frequency values of 
these reflection
coefficients depend on the form of impedance function used.
For the impedance function of the normal and anomalous skin
effect,\cite{56} one reobtains Eq.~(\ref{eq8}) obtained
previously for ideal metals. For the impedance function
of the infrared optics it follows that
\begin{equation}
r_{\rm TM}(0,k_{\bot})=1,
\qquad
r_{\rm TE}(0,k_{\bot})=\frac{\omega_p-
ck_{\bot}}{\omega_p+ck_{\bot}}.
\label{eq17}
\end{equation}
\noindent
In the limit of ideal metal plates 
$\omega_p\to\infty$ and Eq.~(\ref{eq17}) coincides with
Eq.~(\ref{eq8}). The Leontovich impedance leads to almost the
same results for the thermal Casimir force
as the plasma model, i.e., to small thermal
corrections to the zero-temperature force at short separations and
to Eq.~(\ref{eq9}) at large separations.

From the above it is seen, that there are three theoretical
approaches using the Drude model, the plasma model and the 
Leontovich impedance which lead to different predictions for the
thermal Casimir force. There is also the similarity
between the plasma model approach and the impedance
approach which both predict small thermal effects at short
separations and are in agreement with the classical limit
at large separations. This is in opposition to the Drude
model approach which predicts relatively large thermal
effect at short separations and is in violation of the
classical limit at large separations.

As was analytically proved in Refs.~\refcite{59,60} (see
also Refs.~\refcite{25,34}), the Drude model approach leads
to a violation of the third law of thermodynamics (the Nernst
heat theorem) in the case of metallic perfect lattices with
no defects and impurities. For such lattices the relaxation
parameter $\gamma(T)\to 0$ when $T\to 0$ in accordance with
the Bloch-Gr\"{u}neisen law and the entropy of a fluctuating
field at zero temperature takes a negative value\cite{25,34,60}
\begin{equation}
S(z,0)=\frac{k_B}{16\pi z^2}\int\limits_{0}^{\infty}y\,dy\,
\ln\left[1-\left(\frac{cy-\sqrt{4z^2\omega_p^2+c^2y^2}}{cy+
\sqrt{4z^2\omega_p^2+c^2y^2}}\right)^2e^{-y}\right]<0,
\label{eq18}
\end{equation}
\noindent
instead of zero as is demanded by the Nernst heat theorem.
At large separations from Eq.~(\ref{eq18}) it follows
\begin{equation}
S(z,0)=-\frac{k_B\zeta(3)}{16\pi z^2}<0,
\label{eq19}
\end{equation}
\noindent
i.e., what is called in Refs.~\refcite{26,51}--\refcite{53}
the entropy of a ``modified ideal metal'' (MIM) at zero
temperature. Recent Refs.~\refcite{26,53} recognize that
their MIM violates the Nernst heat theorem but argue\cite{26}
that ``the crucial difference between real metals and MIM
is that the former includes relaxation by which there will be
no violation of the third law of thermodynamics''. This
conclusion is wrong because Eq.~(\ref{eq18}) proves the 
violation of the Nernst heat theorem for Drude metals with
dielectric permittivity (\ref{eq10}). These metals have a
finite permittivity at all frequencies with exception of
zero frequency and a nonzero relaxation described by the 
relaxation parameter $\gamma(T)$. From this it follows
that the Drude model approach violates the third law of
thermodynamics for perfect metallic crystal lattices with
no impurities but nonzero relaxation at any nonzero
temperature. Thus, theoretically this approach
is not acceptable.

Several attempts were made to avoid this conclusion. 
In Refs.~\refcite{52,62} the Drude model approach was
applied to metallic lattices with defects and impurities
possessing some residual relaxation $\gamma(0)\neq 0$.
As a result, the equality $S(z,0)=0$ was obtained which is in
accordance with the Nernst heat theorem. This, however, does
not solve the problem of the thermodynamic inconsistency
of the Drude model approach, because metallic perfect crystal
lattice with no impurities has a nondegenerate dynamic state
of lowest energy. Thus, according to quantum statistical
physics, the entropy at $T=0$ must be equal to zero
for such crystal lattices
[a property violated by the Drude model approach
according to Eq.~(\ref{eq18})].

Another attempt\cite{63} includes spatial dispersion in the
calculations of the Casimir energy. At large separations it
arrives at the same Eq.~(\ref{eq12}) as was obtained by
using the Drude model. At arbitrary separations between the
plates computations in Ref.~\refcite{63} nearly exactly
coincide with earlier computations\cite{50} using the
Drude model. In Refs.~\refcite{29,34,64} it was demonstrated,
however, that the results of Ref.~\refcite{63} are not reliable
because the used approximative description of a spatial
dispersion is unjustified. The main mistake in
Ref.~\refcite{63} is that it uses the standard continuity
boundary conditions (\ref{eq6}) 
on the electromagnetic field which are valid only in
the absence of spatial dispersion. If the spatial dispersion
is present, one must use instead the more complicated
conditions\cite{65}
\begin{equation}
E_{1t}=E_{2t}, \quad
B_{1n}=B_{2n}, \quad
D_{2n}-D_{1n}=4\pi\sigma, 
\quad
\left[\mbox{\boldmath$n$}\times({\mbox{\boldmath$B$}}_2-
{\mbox{\boldmath$B$}}_1)\right]=\frac{4\pi}{c}
{\mbox{\boldmath$j$}},
\label{eq20}
\end{equation}
\noindent
where the induced charge and current densities are given by
\begin{equation}
\sigma=\frac{1}{4\pi}\int\limits_{1}^{2}\mbox{div}
\left[\mbox{\boldmath$n$}\times\left[{\mbox{\boldmath$D$}}
\times{\mbox{\boldmath$n$}}\right]\right]\,dl,
\quad
{\mbox{\boldmath$j$}}=\frac{1}{4\pi}\int\limits_{1}^{2}
\frac{\partial{\mbox{\boldmath$D$}}}{\partial t}\,dl.
\label{eq21}
\end{equation}

In the Reply\cite{66} to the Comment\cite{64} the author
attempts to avoid this conclusion by introducing the
auxiliary fields and by bringing the Maxwell 
equations to the form
with no induced charge and current densities.
This attempt, however, fails because, as the author 
himself recognizes, the relations used by him are
valid only in the Fourier space. In the case of
temporal dispersion there is no problem in making
the Fourier transform. However, for spatial dispersion
in the presence of boundaries and a macroscopic gap
between the two plates, this is not allowed.\cite{64}
The system under consideration in the Casimir effect
is not spatially uniform and it is not possible to
introduce the dielectric permittivity
$\varepsilon({\mbox{\boldmath$q$}},\omega)$ depending 
on both the wave vector and the frequency as is done
in Refs.~\refcite{63,66}.

Reply\cite{66} denies the note in the Comment\cite{64}
that the formalism used in the original work\cite{63}
involves nonconservation of energy. In support
of this denial, it is argued
 that the energy leaving a region through an
interface is entering the region on the other side,
and, thus, energy is fully conserved. To arrive at
this conclusion, the author admits that the in-plane
components of the fields are continuous across the
interface. In the presence of spatial dispersion this
assumption is, however, not valid, as was demonstrated above. 
We underline that the violation of energy conservation  
in the so-called ``dielectric approximation'' of
nonlocal electrodynamics used in 
Refs.~\refcite{63,66} has long been
rigorously proved\cite{66a} and discussed in the
literature.\cite{65}

To conclude, presently there is no question that the
approach to the thermal Casimir force using the Drude 
model is thermodynamically invalid. At the same time,
the plasma model and impedance approaches are
consistent with thermodynamics. In particular, they
satisfy the Nernst heat theorem.\cite{25,34}
In Refs.~\refcite{27}--\refcite{29} it was shown
that the same problems, as for metals, arise for
dielectrics if one describes their conductivity
at zero frequency with the help of the Drude
model. This problem is more detailly discussed in
another contribution to these Proceedings.\cite{67}

Important problem is the comparison of different
theoretical approaches to the thermal Casimir force
with experiment. As was already emphasized in Sec.~3,
the computations at zero temperature, and also
theoretical approaches using the plasma model and the 
Leontovich surface impedance at $T=300\,$K, are
consistent  with experiment (see, for example, Fig.~1,
left). At the same time, the theoretical approach
using the Drude model is excluded by experiment
at 95\% confidence level within the separation
region from 170 to 700\,nm. In the separation
region from 300 to 500\,nm the Drude model approach
is excluded by experiment at even higher 99\%
confidence level.\cite{18,19}. For the purposes of
comparison between experiment and theory, the computations 
of the Casimir pressure were done by using the tabulated 
optical data for the complex index of refraction\cite{42}
extended to lower frequencies. In fact, a marked
difference between approaches arises only when 
calculating the contribution of the zero-frequency term in 
the Lifshitz formula which should be found theoretically 
because at very low frequencies 
optical data are
not available.

The comparison between experiment and theory in Fig.~1
is quite transparent. However, in Refs.~\refcite{26,53}
several objections against it were raised. According
to Ref.~\refcite{53}, Purdue group\cite{17,18} claims
``the extraordinary high precision to be able to see our 
effect at distance as small as 100\,nm'' and ``the accuracy
is claimed to be better than 1\% at separations down
to less than 100\,nm''. 
These statements
are misleading because the experimental ranges
in Refs.~\refcite{17} and \refcite{18} are from 260
to 1100\,nm and from 160 to 750\,nm, respectively.
There are no statements concerning 
the separations of 100\,nm and  below 100\,nm 
in Refs.~\refcite{17,18}. According to
another claim in Refs.~\refcite{26,53}, the determination
of the absolute sphere-plate separation with the
absolute error $\Delta z=0.6\,$nm, as stated in
Ref.~\refcite{18}, is difficult because ``the roughness
of the surfaces is much larger than the precision
stated in the determination of the separation''. This
claim is not right because the separations are measured
between zero levels of the surface roughness. These zero
levels are uniquely determined for any value of the
roughness amplitude. One more claim\cite{53}
is that ``the effects of surface plasmons\cite{68,69} have
not been included''. This claim is wrong because the
computations in Refs.~\refcite{17,18} were performed
using the Lifshitz formula which includes in full
the effects of surface plasmons.

As was noticed recently,\cite{70} the precise values of
the Drude parameters are important for an accurate
calculation of the Casimir force in experimental
configurations. According to Ref.~\refcite{70}, the use
of different Drude parameters measured and calculated
for different Au samples may lead to up to 5\%
variations in the magnitude of the Casimir force.
In the computations of Refs.~\refcite{17}--\refcite{19}
the values $\omega_p=9.0\,$eV and 
$\gamma(T=300\,\mbox{K})=0.035\,$eV were used which
are based on the experimental data of Ref.~\refcite{42}
and computations of Ref.~\refcite{71}. As was
demonstrated above, these values lead to a very good
agreement with traditional approaches to the thermal
Casimir force which predict only small thermal
corrections at short separations and exclude the Drude
model approach. If much smaller value for Au plasma
frequency were used in computations (i.e.,
$\omega_p=6.85\,$eV or 7.50\,eV
as suggested in Ref.~\refcite{70}) the agreement between
the traditional theoretical approaches and experimental
data would be worse for a few percent. The same holds
for many other experiments on the Casimir
effect.\cite{13,14,16}${}^{-}$\cite{22}
It should be particularly emphasized that with 
smaller values of $\omega_p$ the disagreement between 
the experimental data and the Drude model approach 
to the thermal Casimir force becomes 
much larger than is demonstrated in Fig.~1 (right).
If one uses widely accepted criteria from the statistical theory of
the verification of alternative hypotheses,\cite{72}
the hypothesis on much smaller magnitude of
$\omega_p$ (than that used in Refs.~\refcite{17}--\refcite{19})
is rejected at high confidence by all already
performed experiments on the Casimir
force with Au surfaces.
This conclusion was recently confirmed\cite{73}
by the determination of the plasma frequency of
Au coatings in the experimental configurations
of Refs.~\refcite{17}--\refcite{19} using the measured
temperature dependence of the films resistivity.
The obtained result 
$\omega_p=8.9\,$eV  [and a respective value for
$\gamma(T=300\,\mbox{K})=0.0357\,$eV] is in excellent
agreement with Refs.~\refcite{42,71}.
It leads to even better than in Fig.~1 agreement of data with
the traditional approaches to the thermal Casimir force
and excludes the Drude model
approach at the impressive 99.9\% confidence level within
a wide separation range. Thus, to date, it is beyond
question that the Drude model approach is experimentally
excluded.

One more important physical phenomenon which sheds light
on the problem of the thermal Casimir force is the modulation
of the Casimir force with laser light discussed in Sec.~3.
From Fig.~3 it follows that the experimental data are 
consistent with theory if the dc conductivity of high
resistivity Si in the absence of laser light is discounted.
On the contrary, the dashed line takes into account dc
conductivity of a Si plate in the absence of laser light
described using the Drude dielectric function.
As is seen in Fig.~3, the dashed line is experimentally
excluded. Thus, for both metals and semiconductors the
account of actual dielectric response at very low
frequencies leads to contradictions between the Lifshitz
theory and the experiment. To achieve an agreement between
experiment and theory, one should use the dielectric
response in the region of characteristic frequency
$\sim c/(2z)$ and extrapolate it to zero frequency.
The complete understanding of this problem goes beyond 
the scope of the Lifshitz theory.

\section{Constraints on new physics beyond the Standard
Model}

Many extensions of the Standard Model predict a new (so-called
``fifth'') force coexisting with the usual Newtonian
gravitational force and other conventional interactions. 
Such force can arise from the exchange of light elementary
particles (e.g., scalar axions, graviphotons, dilatons,
and moduli\cite{74,75}), and as a consequence of
extra-dimensional theories with low energy compactification
scales.\cite{76,76a} The interaction potential of the fifth
force acting between two point masses $m_1$ and $m_2$
at a distance $r$ is conventionally represented as a 
Yukawa correction to the usual Newtonian potential
\begin{equation}
V(r)=-\frac{Gm_1m_2}{r}
\left(1+\alpha e^{-r/\lambda}\right),
\label{eq22}
\end{equation}
\noindent
where $G$ is the gravitational constant, $\alpha$ is
a dimensionless constant characterizing the strength of
the Yukawa force, and $\lambda$ is its interaction range.

It has been 
known\cite{17}${}^{-}$\cite{19,30,77}${}^{-}$\cite{80} 
that the best constraints on the parameters 
$(\alpha,\lambda)$ in submicron range follow from the
measurements of the Casimir force. The pressure
of a hypothetical
interaction, $P^{\rm hyp}(z)$, which may act between
experimental test bodies is computed\cite{17}${}^{-}$\cite{19}
by the pairwise summation of potentials (\ref{eq22}) with
a subsequent negative differentiation with respect to
separation. Then constraints on the hypothetical Yukawa-type
pressure are found from the agreement between measurements and
theory at 95\% confidence level. According to the experimental
results in Refs.~\refcite{18,19}, no deviations from
calculations using the traditional theories of the Casimir
force were observed. Thus, one can conclude that the 
hypothetical pressure should be less than or equal to
the half-width of the confidence interval
\begin{equation}
|P^{\rm hyp}(z)|\leq\Delta^{\!\rm tot}\left[P^{\rm th}(z)-
P^{\rm exp}(z)\right],
\label{eq23}
\end{equation}
\noindent
where $\Delta^{\!\rm tot}\left[P^{\rm th}(z)-
P^{\rm exp}(z)\right]$ is the total absolute error of
the quantity $P^{\rm th}(z)-P^{\rm exp}(z)$. Note that just
this error (and negative this error) are plotted in Fig.~1
(left and right) by the solid lines.
\begin{figure}[t]
\vspace*{-6.5cm}
\hspace*{-1.6cm}
\psfig{file=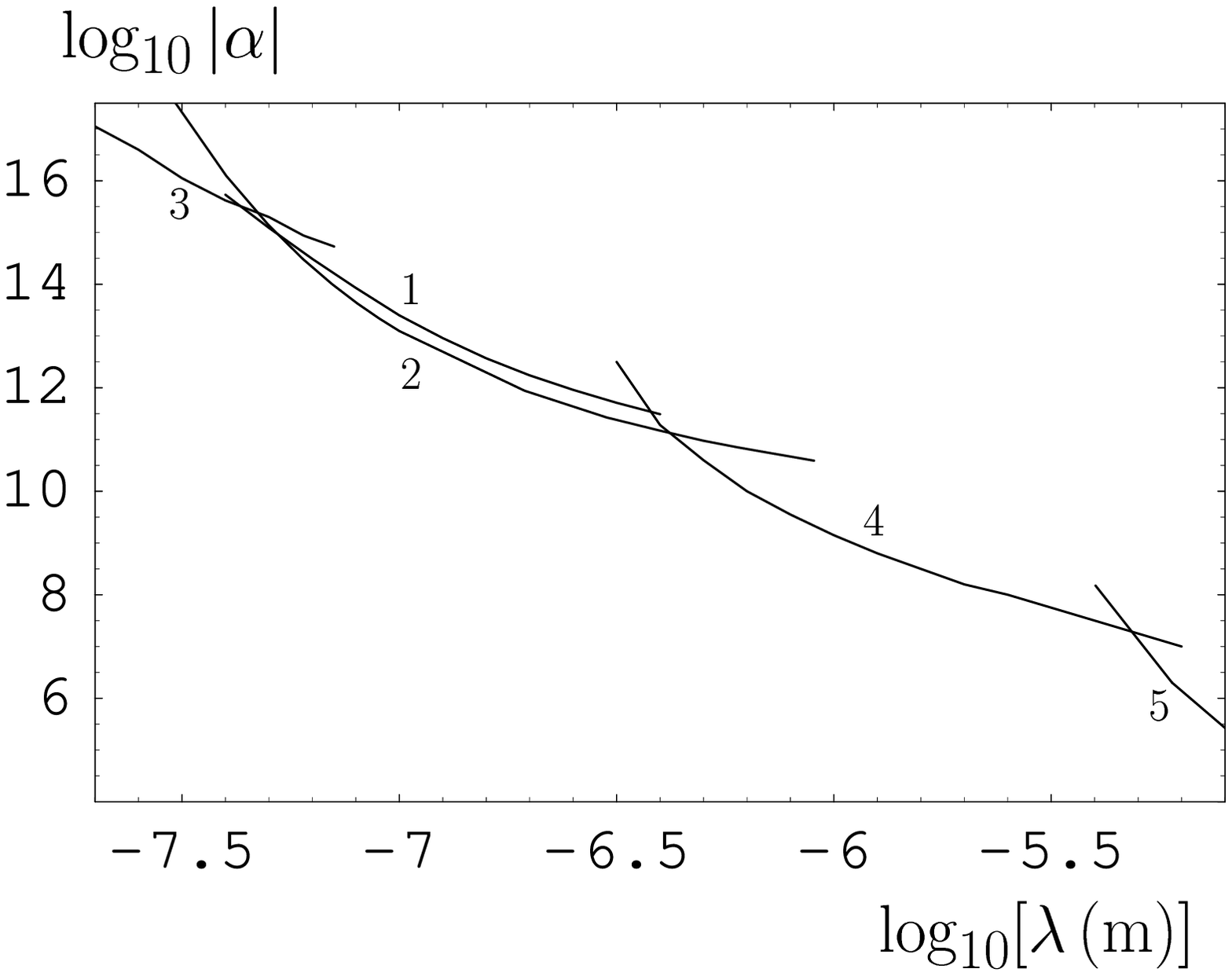,width=6in}
\vspace*{-6cm}
\caption{Constraints on the Yukawa interaction constant
$\alpha$ versus interaction range $\lambda$. Line 1 is
obtained from the measurements of the Casimir preesure
by use of a micromechanical torsional
oscillator.\cite{18,19} Line 2 follows from the isoelectronic
differential force measurements.\cite{30} Line 3 is
obtained from the measurement of the Casimir force using
an atomic force microscope,\cite{13} and line 4 from
the torsion-pendulum experiment.\cite{9}
The strongest constraints following from the gravitational
measurements using a micromachined cantilever\cite{81}
are indicated by the line 5.}
\end{figure}

In Fig.~4 we plot the strongest constraints on $\alpha$
for different values of $\lambda$ following from the
measurements of the Casimir force and compare them with
the best gravitational experiments. Each line in Fig.~4
is related to some specific experiment. The region of
($\alpha,\lambda$) plane above each line is prohibited
from the respective experiment and below each line is
permitted. Constraints
shown by line 1 follow from the most recent
measurement of the Casimir pressure using a
micromechanical torsional oscillator.\cite{18,19}
Line 3 is obtained\cite{80} from the Casimir force
measurement between an Au-coated sphere and a plate
by means of an atomic force microscope.\cite{13}
Line 4 follows\cite{77} from the measurement of the
Casimir force between an Au-coated spherical lens
and a plate by means of torsion pendulum.\cite{9}
Line 2 presents the constraints obtained from the
first isoelectronic differential force
measurement\cite{30} between an Au-coated probe and two
Au-coated films, made out of gold and germanium.
In this measurement the Casimir background is
experimentally subtracted, thus avoiding the
necessity to model the Casimir force. Finally, line 5
shows the most strong constraints on Yukawa-type
deviations from Newtonian gravity obtained\cite{81}
from the gravitational experiment using a micromachined 
silicon cantilever as the force sensor at separations of
order $25\,\mu$m, where the Casimir force is already
negligibly small. Gravitational experiments provide
also the strongest constraints on $\alpha$ in the
interaction range $\lambda>10^{-5}\,$m (see 
Refs.~\refcite{19,74} and \refcite{82} for more details).

As is seen in Fig.~4, at and  below a micrometer interaction
range there is no competitors to the Casimir effect in
obtaining constraints on non-Newtonian gravity.
During the last few years these constraints were
strengthened by up to $10^4$ times basing on the results
of different measurements of the Casimir force between
metal surfaces. Nevertheless, existing limits on
$\alpha$ below a micrometer are still relatively
weak and should be strengthened by several orders of
magnitude to reach the theoretically predicted regions
of strange and gluon modulus, and of gauged barions. 

\section{Conclusions and discussion}

In the foregoing, we have discussed main achievements in
the physics of the Casimir effect during the period after the
Xth Marcel Grossmann Meeting hold in 2003. In our opinion,
the major theoretical breakthrough is the obtaining of
the exact analytical solution for the 
electromagnetic Casimir energy in the
configuration of an ideal metal cylinder above a
plate. The resolution of this problem opened new
opportunities for the investigation of the Casimir force
between curved boundaries and permitted find first
indisputable result on the accuracy of the PFT.

The major experimental breakthroughs during this period are
the most precise measurements of the Casimir pressure between
metal surfaces using a micromechanical 
oscillator (Decca et al.) and the first experiments on the
Casimir effect in configuration of metal sphere
and semiconductor plate by means of an atomic force
microscope (Mohideen et al.). Both sets of experiments
resulted in far-reaching and important conclusions.
Experiments by Decca et al. have conclusively excluded
large thermal corrections to the Casimir force at short
separations as predicted by the Drude model approach.
Experiments by Mohideen et al. demonstrated the possibility
to control the Casimir force by changing the density of
free charge carriers and led to new knowledge on the applicability
of the Lifshitz theory to dielectrics.

Time between Xth and XIth Marcel Grossmann Meetings was
marked by controversial discussions of different approaches to the
theoretical description of the thermal Casimir force. During these
discussions it was clearly demonstrated that all the proposed
approaches are of approximate phenomenological character. None
of them can yet claim to be the final fundamental resolution of the problem.
Till the end of this period it was conclusively demonstrated that
the Drude model approach is in contradiction with the foundations of
thermodynamics and is excluded experimentally at a 99.9\%
confidence level. Important theoretical problem for future is
the fundamental understanding of the thermal Casimir force and 
related physical phenomena caused by vacuum and thermal oscillations
of the electromagnetic field (e.g., atomic friction, radiative
heat transfer etc.). The experimental challenge for near future is
the measurement of the thermal effect in the Casimir force
which has not been measured yet.

Both recent theoretical achievements and the performed experiments
confirm the unique potential of the Casimir effect both in
fundamental physics for constraining predictions of new 
unification physical theories beyond the Standard Model and
in nanotechnology for fabrication, operation and control of a
new generation of microdevices. This confirms the increasing role of the
Casimir effect both in modern physics and in technological
applications.

\section*{Acknowledgments}

The author is grateful for helpful discussions with
M.~Bordag, R.~S.~Decca, E.~Fischbach, B.~Geyer,
G.~L.~Klimchitskaya, D.~E.~Krause, and U.~Mohideen.


\end{document}